# Transferable Molecular Charge Assignment Using Deep Neural Networks

Ben Nebgen,[a] Nick Lubbers,[a] Justin S. Smith,[a,b] Andrew Sifain,[a,c] Andrey Lokhov,[a] Olexandr Isayev,[d] Adrian Roitberg,[b] Kipton Barros,[a] Sergei Tretiak[a]

We use HIP-NN, a neural network architecture that excels at predicting molecular energies, to predict atomic charges. The charge predictions are accurate over a wide range of molecules (both small and large) and for a diverse set of charge assignment schemes. To demonstrate the power of charge prediction on non-equilibrium geometries, we use HIP-NN to generate IR spectra from dynamical trajectories on a variety of molecules. The results are in good agreement with reference IR spectra produced by traditional theoretical methods. Critically, for this application, HIP-NN charge predictions are about $10^4$ times faster than direct DFT charge calculations. Thus, ML provides a pathway to greatly increase the range of feasible simulations while retaining quantum-level accuracy. In summary, our results provide further evidence that machine learning can replicate high-level quantum calculations at a tiny fraction of the computational cost.

## Introduction

The solution to the Schrödinger equation of quantum mechanics (QM) provides, in principle, a complete description of all chemical phenomena. However, exact solutions are infeasible for systems of practical interest due to the exponential scaling of computational cost with molecular size (i.e., the number of constituent atoms). Thus, modern theoretical chemistry has turned to a wide range of approximate methods for the simulation of chemical systems with an important trade-off between accuracy and computational cost. For example, one of the most popular theoretical methods (with over 15,000 publications per year)[1,2] is density functional theory (DFT), where the computational effort is reduced to roughly cubic scaling with molecular size. Yet, even cubic scaling is prohibitive for many applications. Physically relevant length and time scales are often completely inaccessible to DFT and higher order QM methods. The development of more efficient methods that can retain quantum accuracy will open the door to a wealth of useful studies for physics, chemistry, biophysics, and materials science.

A conventional approach to address these challenges is the formulation of better numerical techniques, e.g., to enable more efficient solution of DFT and related theories. An alternative and emerging approach is to use Machine Learning (ML) models, which are poised to revolutionize the state of electronic structure methods. In particular, ML may offer the accuracy of high fidelity quantum mechanical calculations at a fraction of cost. ML methods commonly scale linearly with the number of atoms.[3,4] A typical ML approach utilizes a reference dataset and a learning algorithm which constructs a robust model of the data. Because it is data-dependent, ML is not a replacement but a complement to existing electronic structure theory enabling multi-step modelling. Reference calculations are typically used to build a database of small molecules or fragments which represent "building blocks" for extended systems. Such databases may be strengthened by a subset of high quality experimental data and/or high order QM simulations. For example, prediction of molecular energies using ML has been explosively developing in recent years.[4-13] Modern ML models commonly fit DFT energy datasets to within 0.5 kcal/mol (1 kcal/mol is frequently quoted as a gold standard for chemical energy accuracy). Remarkably, this accuracy far surpasses the precision of DFT itself. Moreover, once trained, a ML prediction is usually thousands of times faster than a conventional DFT calculation. There are many other applications beyond energy prediction for ML in quantum chemistry; reported examples include ML of band-gap calculations,[14,15] high throughput screening for materials discovery,[16,17] crystal structure prediction,[18] and excited state dynamics.[19-21]

[a] Theoretical Division, Los Alamos National Laboratory, Los Alamos, New Mexico 87545
[b] Department of Chemistry, University of Florida, Gainesville, Florida 32611
[c] Department of Physics and Astronomy, University of Southern California, Los Angeles, California 90037
[d] UNC Eshelman School of Pharmacy, University of North Carolina Chapel Hill, Chapel Hill, North Carolina, 27599

Atomic charges constitute an important quantity allowing to rationalize and extend our chemical intuition, and to calculate a variety of molecular properties. All charge assignment schemes require complete quantum-mechanical calculation of the electronic density with subsequent partitioning of the latter into atom-centered point charges.[22] These steps add additional numerical complexity on top of the electronic structure calculation. Such simulations are critical to a broad area of theoretical chemistry, ranging from modelling of infrared (IR) spectra,[23] evaluation of solvation free energies,[24,25] classical force field parameterization,[26,27] etc. Recent work used ML to predict IR spectra of a single system at a time.[28] Parkhill and co-workers have proposed a general neural network model[3] able to predict energies, forces, and charges simultaneously, which can be used for modelling of IR spectra. Yet, many questions remain unexplored, including *extensibility* (i.e., how well can we predict charges on systems much larger than those included in the training set?), *charge partitioning reliability* (i.e., can all charge assignment methods be well learned?), and *transferability* (i.e., can learning from datasets of small but diverse sets of molecules be applicable to larger systems?).

Here, we use the HIP-NN[5] (Hierarchically Interacting Particle Neural Network) ML model to predict atomic charges under a variety of charge assignment schemes. Our goals are three-fold: First, we demonstrate that HIP-NN can accurately and efficiently predict charges for a wide range of small molecules, which is critical for classical force field parameterization. Second, we show that HIP-NN can effectively predict many different charge assignment schemes. Finally, we validate the extensibility of ML models by training them to databases of small organic molecules and applying them to significantly larger systems such as drug and peptides.

**Methods**

**Charge Partitioning Schemes**

From a QM perspective, charge is distributed as a continuous density field across the entire molecule. Unfortunately, there is no unique or exact way to condense this field onto individual constituent atoms. There are three broad classes of charge assignment schemes that can be derived from *ab-initio* simulations.[22] First, the electron density obtained from the *ab-initio* calculations can be partitioned directly onto each atom. In this category, Mulliken charges[29] are the simplest, cheapest, and least accurate. Hirshfeld charges[30] are more sophisticated and accurate. Finally, Natural Bond Orbital (NBO)[31,32] package assigns charges according to the Natural Atomic Orbitals of a molecule, which are orbitals designed to capture the behaviour of atomic orbitals in a given molecular environment. The second family of approaches is to use *ab-initio* calculations to produce molecular properties, which guide charge assignment. Popular choices include charge multipoles and the electrostatic field at specific points surrounding the molecule. Then, the charges are fit to best recover these properties. Merz-Singh-Kollman

(MSK)[33] charges are fit to reproduce both the molecular dipole and charge distributions on large, complex molecules used in classical force-fields. In the last family, one combines *ab-initio* electronic density with gas-phase experimental measurements using a simple correction. An example is Charge Model 5 (CM5)[22], in which the correction is parameterized to typically replicate dipole moments for electrostatic energy calculations. Unlike MSK charges, CM5 charges are not constrained to exactly replicate the quantum molecular dipole. Rather, one expects CM5 charges to produce an approximately correct dipole moment. Section 1 in the Electronic Supporting Information (ESI) contains further details on these charge schemes. In this work, we test our ML techniques against commonly used Hirshfeld, NBO, MSK, and CM5 charge schemes.

To illustrate these distinct charge schemes and to establish a baseline accuracy for our ML model, we consider charge assignment in a methanol molecule ($CH_3OH$). We select Methanol because it has a strong molecular dipole moment and is simple enough for chemical intuition to be applicable. In Table 1, we report the charge on the carbon and oxygen atoms in methanol on a variety of charge schemes. The charges were generated using DFT calculations with wB97x functional and two different basis set sizes (additionally, results obtained with a simpler 6-31G basis are reported in the ESI). There is a significant variation between the methods. Of the four charge schemes, only MSK assigns a positive charge to the carbon atom. The Hirshfeld method attributes a nearly neutral charge to the carbon, while the remaining schemes assign various negative charge values to this atom. This demonstrates how differing charge schemes may operate in fundamentally disparate ways, each posing an independent challenge for ML modelling.

Both CM5 and Hirshfeld charges are nearly independent of basis set, with the difference between assigned charges under .01 $e^-$. NBO has a larger basis set dependence, with the charge variation within .03 $e^-$. Here, the local atomic charge densities that are used to partition the molecular charge density are nearly independent of basis set size. In contrast, MSK charges vary greatly with the basis set size. This is because the global constraint to reproduce the dipole moment introduces non-locality into the charge partitioning.

It is not obvious whether larger basis sets result in more practical charge assignments.[34] As basis sets get larger and orbitals are more diffuse, associating charge with specific atoms becomes more ambiguous. Within a single charge assignment scheme, it is unreasonable to expect errors smaller than 0.01 $e^-$. Thus, in the following analysis, we aim to build ML models of charge assignment that are accurate to within 0.01 $e^-$.

| Atom | Basis set | Hirshfeld | CM5 | MSK | NBO |
| --- | --- | --- | --- | --- | --- |
| Oxygen: | 6-31G* | -0.255 | -0.475 | -0.618 | -0.751 |
| Oxygen: | 6-31+G* | -0.252 | -0.472 | -0.699 | -0.777 |
| Carbon: | 6-31G* | -0.007 | -0.132 | 0.142 | -0.317 |
| Carbon: | 6-31+G* | -0.010 | -0.134 | 0.234 | -0.334 |

Table 1: Charges on carbon and oxygen atoms in in methanol according to various charge partitioning schemes and typical basis sets.

**HIP-NN approach**

Here we briefly describe Deep Neural Networks for chemical property prediction,[35, 36] and, in particular, the Hierarchical Interacting Particle Neural Network (HIP-NN) used in this work. For details, see Section 2 in the ESI and Ref [5].

To begin, a molecule is converted into a feature representation that can be parsed by a neural network. A variety of schemes are available,[37-42] such as the Coulomb matrix[43] or atom centred symmetry functions suggested by Behler and used in ANI-1.[4, 11] The features used for HIP-NN are relatively minimal: We use the atomic number of each atom, and the pairwise distances between atoms. This simple representation ensures the network predictions satisfy translational, rotational, and reflection invariances. The features are passed through many *hidden layers* to produce a vector of new internal features or *activations* at each layer. Each layer's activations are produced using matrix-vector products and an element-wise nonlinearity. Finally, *output layers* form linear combinations of the activations to produce predictions at each atom. The internal weights used in these computations constitute the learnable *parameters* of the network. These parameters are fit to match the training dataset using an iterative optimization process. In this work, each network has about $10^4$ parameters.

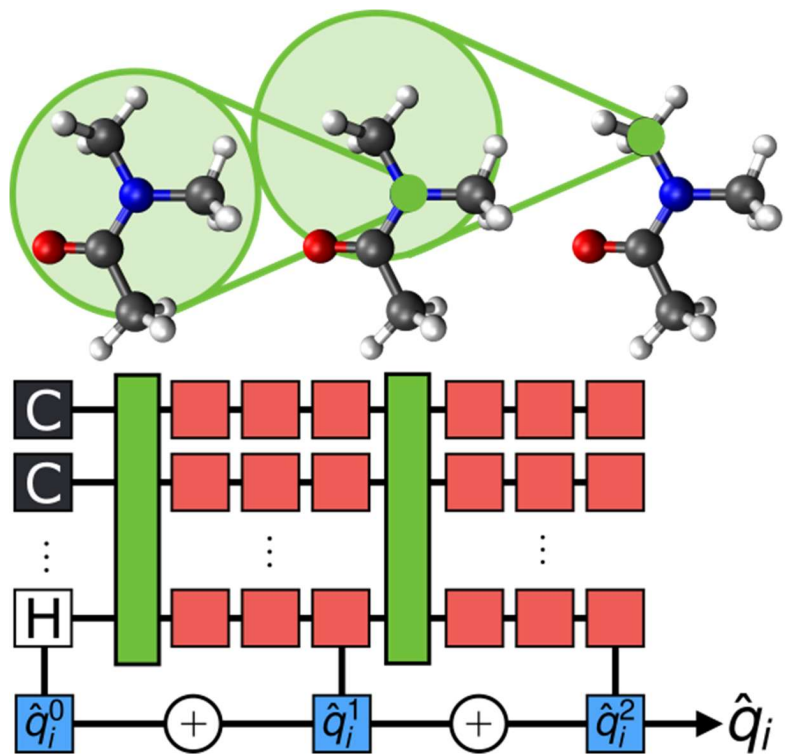

Figure 1: A diagrammatic representation of HIP-NN. Green bars represent interaction layers and red squares represent on-site layers for atoms. The blue squares are the output layers which return a series of corrections to the atomic charge.

The structure of HIP-NN is depicted in Fig. 1. The network variables reside on each atom in the molecule, whose initial features encode the species of the atom. Green bars illustrate interaction layers, which allow sharing of information between nearby atoms. Interaction layers are comprised of a number of sensitivity functions, which allow atoms at different distances to interact in different ways. These sensitivity functions are constrained by two cut-off distances: The soft cut-off is the distance at which all interactions begin decreasing, and the hard cut-off is when all interactions fall to zero. Red squares in Fig. 1 illustrate on-site layers, which perform processing of activations on each atom independently. Additionally HIP-NN is organized into *interaction blocks*, each of which contains a single interaction layer, followed by a number of on-site layers. At the end of each series of on-site layers, an output layer gathers the current information on a given atom and produces a contribution to the atomic charge. In this way, the total charge on an atom is given by a sum over layer-wise contributions:

$$\hat{q}_i = \sum_{j=0}^{N_{interaction\_layers}} \hat{q}_i^j$$

We use ADAM[44], a variant of stochastic gradient descent (SGD) to learn the parameters for our networks. Briefly, the source dataset is randomly partitioned into sets: training (60%), validation (20%), and test (20%). The training data is repeatedly fed through the network, and the parameters of the network are adjusted so the outputs of the network tend towards the true reference charges (i.e., those from the given charge assignment). After each pass of the training dataset, the performance on the validation set is recorded in order to estimate how the current parameters generalize to new data. The training process is terminated once the performance on the validation set stops improving. This early stopping procedure helps to prevent *overfitting* (i.e., helps the network to learn generalizable patterns, rather than irrelevant details specific to the training set). The model that scored the best on the validation set is kept, and is used to measure the out-of-sample performance on the test set. Because the test set does not play any role in the fitting process, the error on the test set constitutes a fair measure of the network performance on data that is similar in scope to the training dataset. Section 2 of the ESI contains a more detailed description of the training process.

**Training Databases and Tested Networks**

Here our choices were motivated by previous experience of learning energies within HIP-NN and ANI-1[4] approaches. Two different databases were used to train HIP-NN in an effort to understand the transferability of ML algorithms. The first we denote *GDB-5*: A subsample of GDB-11[45, 46] dataset used to train ANI-1 ML force field.[4] The configurations include up to 5 heavy atoms (of types C, N, and O). The conformations are based on the ANI-1 normal mode sampling scheme. This dataset contains 517,133 individual molecular structures. The subsampling of the GDB-11 was done for two reasons: First, the smaller dataset allowed us to produce charges for many different partitioning schemes without expending excessive computational resources on DFT calculations. Second, the dataset of only small structures is used to determine how networks trained to small molecules can predict charges on larger systems. The Gaussian09[47] computational suite was used to run the DFT calculations under the wB97x/6-31G*[48, 49] functional and basis set. For each molecule in the dataset, Hirshfeld, CM5, NBO, and MSK charges were generated for every atom. This allowed us to construct networks trained to reproduce each type of charge assignment.

The second database examined, denoted *ANI-1x*, was developed by Smith et al.[50] using an Active-Learning method, whereby the training set is progressively built by intelligently performing new DFT simulations to cover the regions of greatest model uncertainty.[4, 28, 51-54] While a summary of method can be found in Section 3 of the ESI, one key component of this procedure is "Query-by-Committee".[55] By training an ensemble of neural networks to identical data and using them to predict values on the same unknown system, the variation between the predictions becomes a proxy for the confidence of that prediction. That is to say, if the particular chemical system is well represented in the training data, all the networks will give the same prediction, while if the system is not well represented in the training data the predictions of the various networks will be widely disparate. The complete technical details are described in Ref. [50]. ANI-1x was produced using a sophisticated method for sampling chemical space and only performing reference calculations on those systems for which an ensemble of neural networks produced no consensus energy value. The procedure produced a dataset of 5.5 million structures. For our charge modeling, we use 6% of ANI-1x for the training set, and an additional 2% each for test and validation sets.

Training a single HIP-NN model to the GDB-5 energy dataset takes roughly 12 hours with GPU acceleration using a NVIDIA 1080Ti, while training to ANI-1x takes roughly 120 hours. Roughly a factor of 3 can be saved by initially training on GDB-5 and then "transferring" the model to train on ANI-1x database. We observe a similar cost savings for HIP-NN charge models trained using the same transfer learning protocol. No significant loss of accuracy was observed between the traditionally trained network and the transfer network, as seen in Table S1 of the supporting information.

Table 2 describes the various forms of HIP-NN that are examined in this work. The R=6.0 network was trained on the small GDB-5 training data. The goal of this network is to allow all distance functions to be trained even when using datasets of very small molecules. The R=4.0 network is a network designed to focus only on nearest neighbour interactions. The R=8.0 network closely resembles the design for predicting energy in the original HIP-NN paper.[5] For each network, a reasonable number of sensitivity functions was chosen based on the cut-off distances to prevent having too many free parameters.

## Results

### Benchmark Databases

To illustrate the generality of HIP-NN, we computed Mulliken, Hirshfeld, CM5, MSK, and NBO reference charges for the entire GDB-5 database and trained the same HIP-NN network on each of the charge schemes. Figure 2 reports the mean absolute error (MAE) and root mean squared error (RMSE) for each network trained to its respective charge scheme. An RMSE much greater than MAE indicates that distribution of errors has a long tail, i.e., it contains outliers. HIP-NN learns almost all of the charge schemes with a MAE of about .005 $e^-$ and a slightly larger RMSE. This is significantly better than our target precision of 0.01 $e^-$, set by the consistency of the charge models. Additionally, the prediction of these charges takes approximately 0.234 ms per conformation. Thus, using these neural networks to predict charges on molecules similar to those found in GDB-5 is significantly computationally cheaper computationally, but no less precise, than using a quantum calculation. The one exception to this is MSK charges, which are replicated to approximately a precision of 0.02 $e^-$. We believe the explanation is as follows: MSK charges are constrained *post-hoc* to exactly replicate the dipole moment of the molecule. This makes the charge scheme non-local; charges may differ between similar conformations in order to produce the correct dipole moment. This non-locality is potentially difficult to represent within the HIP-NN model, which is local.

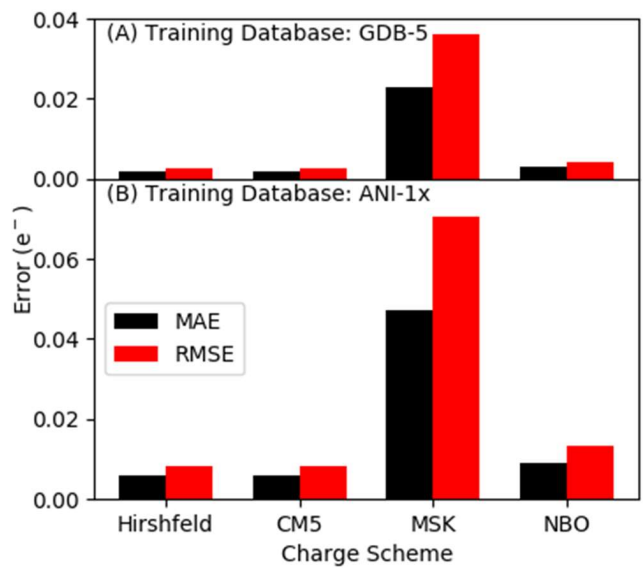

Figure 2: Test set mean absolute and root-mean-square errors for the R=6.0 (see table 5) network trained to the GDB-5 database (A) and ANI-1x database (B) using different charge schemes. HIP-NN is able to learn almost all charge schemes to equal precision, with the exception of the MSK scheme. While the test set error for the more diverse ANI-1x dataset is larger than the test set error for the GDB-5 dataset, the predictive accuracy of networks trained to ANI-1x is significantly better (see Fig. 3)

To test the transferability and extensibility of networks trained to both the GDB-5 and ANI-1x databases, we applied various versions of HIP-NN (Table 2) to the COMP6 benchmark suite[50] of organic molecules (Fig. 3). COMP6 samples a diverse selection of molecular configurations and conformations aimed at validating the accuracy of an ML potential. This suite supplies energies, forces, and Hirshfeld and CM5 charges for the validation of ML methods, all computed using wB97x/6-31G*. COMP6 contains six benchmarks, each containing different types of molecules: GDB-7-9 (built from GDB-11[45, 46]), GDB-10-13 (built from GDB-11[45, 46] and GDB-13[56]), Tripeptides, Drugbank,[57] ANI-MD, and S66x8 non-covalent interaction[58] benchmarks. These benchmarks are described below:

The GDB-7-9 and GDB-10-13 benchmarks aim to validate the universality of the predictor on a comprehensive list of artificially generated small molecules with many non-equilibrium conformations per molecule. These sets contain 1500 and 2996 configurations with 36,000 and 47,670 total conformers, respectively. The tripeptide benchmark provides 1984 random configurations for 248 generated tripeptides. The Druglike benchmark provides 13,379 random configurations of 837 drug molecules from the DrugBank[57] database. ANI-MD contains random structures from MD trajectories generated with the ANI-6a active learned ML potential[50] for 13 common drug molecules and a small protein. The S66x8[58] is a benchmark of 66 different interacting dimers for validating a methods accuracy in reproducing non-covalent interactions. Interactions included in this dataset include hydrogen bonding, London interactions, and π–π stacking.

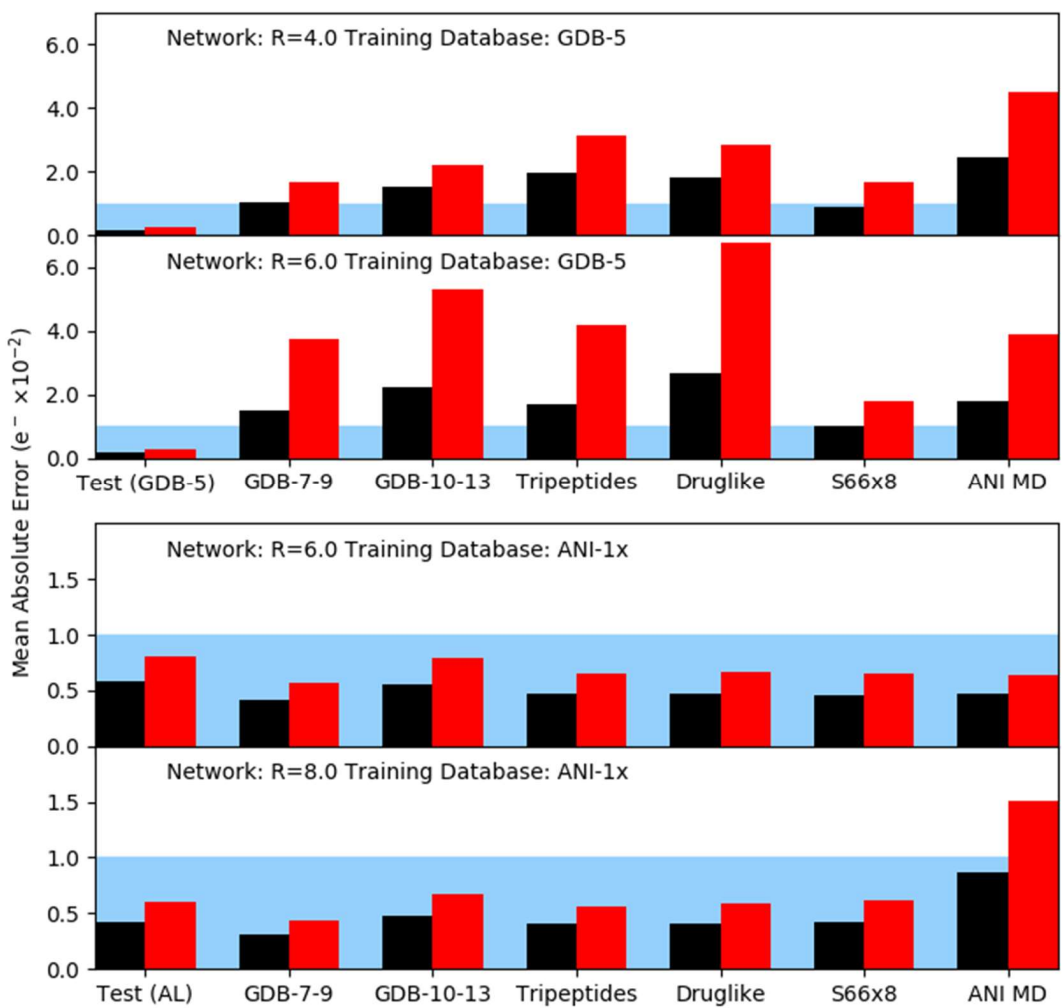

Figure 3: Predictions of various versions of HIP-NN for extrapolation. All charges are from the Hirshfeld partitioning scheme. Mean absolute error (MAE) and root mean squared error (RMSE) given for all datasets. The Blue shading indicates 0.01e$^-$, the target error for this method. A full table of this data can be found in the ESI.

Figure 3, column 1, shows accuracies of the networks given in Table 2 on their respective charge schemes and test sets. It is clear that the GDB-5 charges are much easier to learn than those in ANI-1x. This is due to the smaller size of GDB-5, as well as the lower variety in the data. For example, GDB-5 lacks non-bonding interactions. The rest of the columns in Fig. 3 demonstrate the network performance on the benchmarks in COMP6. The networks trained to GDB-5 predict charges on the benchmark data with significantly lower accuracy; training to GDB-5 does not generalize well. Still, the R=6.0 network trained on the smaller dataset predicts charges with an accuracy of roughly 0.02 $e^-$ for the larger molecules. This error is less than the difference in assigned charge between the two basis sets observed in Table 1 for all charge assignment schemes except Hirshfeld. Thus, it is reasonable to train such a short-range network to a small dataset and still reproduce charges within their margin of error. Section 4 of the ESI contains the exact training accuracies across all benchmarks.

Predictions on the COMP6 dataset made by networks trained to the ANI-1x dataset have a much lower MAE and RMSE than those trained to GDB-5. For example, the MAE (0.0047 $e^-$) of predictions made by the R=6.0 network trained to the ANI-1x database is less than the dependence on the basis set size for even Hirshfeld charges (0.0048 $e^-$ for carbon). This is to say that the accuracy of the predictions is greater than the certainty of the charge assignment itself. However, the dataset itself is harder to learn, with test set errors roughly 3 times larger than those obtained when training to GDB-5 (Fig. 3, Test column).

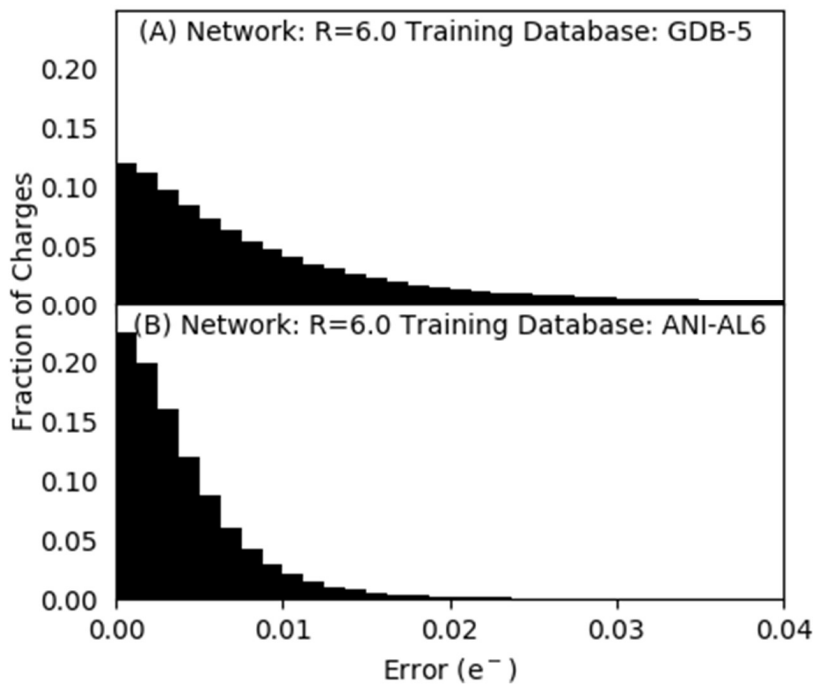

Figure 4: Histogram of errors for the R=6.0 network (see Table 3) when applied to the GDB-7-9 database. The networked trained to ANI-1x (Frame B) database has a significantly shorter tail than the network trained to the GDB-5 database (Frame A).

Figure 4 displays the error distribution of two R=6.0 networks – one trained to GDB-5, and one to ANI-1x – each applied to the GDB-7-9 benchmark. The GDB-5 network (Fig. 4A) error distribution demonstrates the significantly worse performance compared to the ANI-1x network (Fig. 4B). The GDB-5 error distribution has an elongated tail, while the ANI-1x error distribution is roughly normal.

In general, we conclude that accurate and extensible charge prediction can be obtained by learning to a dataset that contains larger molecules and more data diversity, even though such datasets are more difficult to learn, producing a larger MAE test set error. Once networks trained to GDB-5 and ANI-1x are applied to new sets of molecules, it quickly becomes apparent that those models trained to the more diverse dataset outperform those trained to the smaller dataset.

| Network Name | Layer 1 | | | | Layer 2 | | | |
|---|---|---|---|---|---|---|---|---|
| | Sensitivity Functions | Soft Cut-off (Å) | Hard Cut-off (Å) | Number of Units | Sensitivity Functions | Soft Cut-off (Å) | Hard Cut-off (Å) | Number of Units |
| R=4.0 | 8 | 2.5 | 4.0 | 20 | 8 | 2.5 | 4.0 | 20 |
| R=6.0 | 10 | 3.0 | 6.0 | 20 | 8 | 2.5 | 4.0 | 20 |
| R=8.0 | 20 | 5.5 | 8.0 | 40 | 20 | 5.5 | 8.0 | 40 |

Table 2: Parameterization for various networks applied to charge training. Sensitivity Functions shows the number of unique functions that facilitate the interactions between atoms at each interaction layer. Cut-off is the largest possible distance to the maximum of the last spatial sensitivity function and hard cut-off is when all spatial sensitivity functions go to zero. Each network has three atomic layers following each interaction layer. Networks are named for the hard cut-off of the first interaction layer. As each network has two interaction layers, the maximum possible range (receptive field) of the network prediction is the sum of the cut-offs for the interaction layers.

**Simulation of IR Spectra**

One of the applications for a rapid charge prediction scheme is the construction of IR spectra from molecular dynamics data. Here we construct fully ML based IR spectra by generating gas phase MD trajectories for a few small molecules (methanol, ethanal, acetamide and dimethylacetamide) using the ANI-1x potential.[50] Trajectories of 100ps with a 0.1 fs time step were collected on a single molecule at 300K using an NVT thermostat at 300K. From the trajectory conformations, Neural Network charges were generated at every point using the R=6.0 HIP-NN trained to GDB-5 Hirshfeld charges. GDB-5 serves as a sufficient training database because the molecules in the IR study are of a similar size and composition. Using the position and ML charge data for each frame, a molecular dipole is constructed. The IR spectrum is constructed from Fourier transform of the autocorrelation of the dipole moment using the code by Efrem Braun.[59] To produce a fair comparison to DFT, we also generate the IR spectrum with the same algorithm, but using true Hirshfeld charges computed from the same trajectory. We refer to these as the "ML charge IR spectra" and the "QM charge IR spectra," respectively.

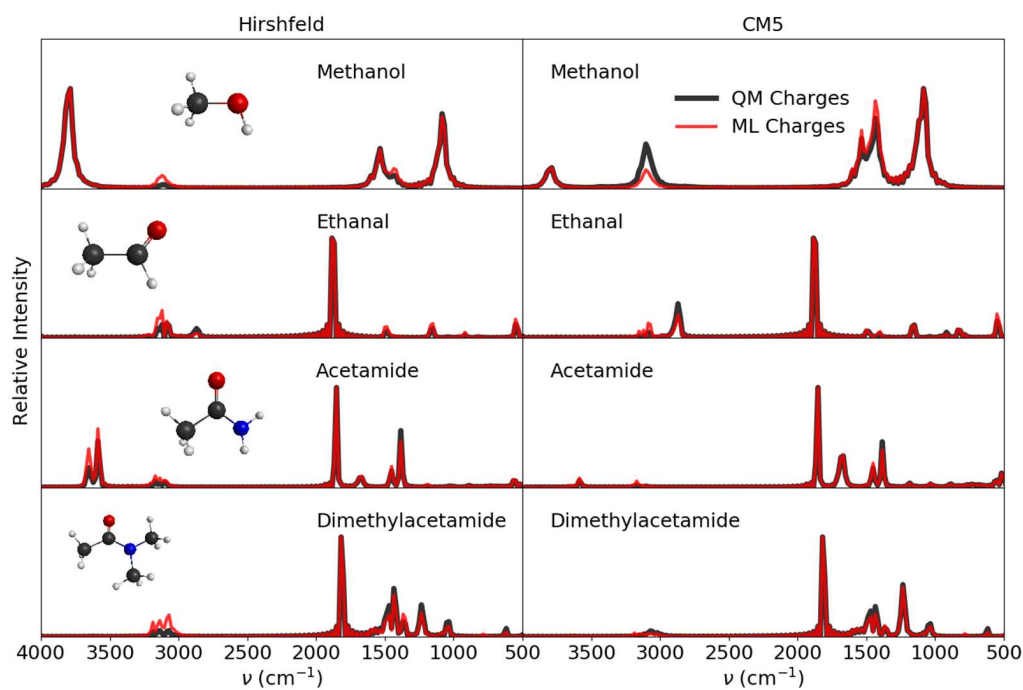

Figure 5: IR spectrum generated completely by machine learning (red line) and from a quantum mechanics/machine learning hybrid (black line). For both spectra, a 100 ps MD trajectory was performed with ANI-1x forces. At each point in this trajectory Hirshfeld charges were assigned to each atom, either by HIP-NN (red line) or from a DFT calculation (black line). The dipole from these charge assignments was then used to construct the IR spectrum in both cases.

Figure 5 shows the ML charge IR spectra in comparison to the QM charge IR spectra. For all molecules, the agreement is excellent. The only exception is the methanol C—H stretch for the CM5 charge spectrum at about 3000 cm$^{-1}$, which is underestimated. The excellent agreement between ML and QM spectra demonstrates that the ML prediction scheme correctly represents charge dynamics in these molecules. See Section 5 of the ESI for comparisons against harmonic normal mode spectra and experimental spectra.

Computationally, charge generation using HIP-NN is extremely fast, requiring roughly 5 minutes on a GPU-equipped workstation. For comparison, the analogous quantum calculations on every point of the molecular trajectory required between 2000 (ethanal) and 3000 (dimethylacetamide) CPU hours, leading to a wall-time speedup of 4 orders of magnitude.

**Charge Prediction on Large Molecules**

Another application of our models is the prediction of charges on systems too large for quantum chemistry, e.g. proteins. Due to the speed of ML charge prediction, it is possible to dynamically assign charges during a MD run for an entire protein, allowing for a dynamic estimate of the coulomb contribution to the force field. As a proof of concept, we have applied the above charge models to glucoamylase protein (1AYX), which has a mass of over 50kDa, and a length of 492 residues.[60] Like our training database, it contains only H, C, N, and O atoms. For prediction, crystallographic water was removed, and protons were added using the Reduce program.[61] We have made the full list of charge assignments in the ESI for Hirshfeld, CM5, NBO, and MSK charges. In total, this methodology took less than 2 minutes to predict charges on the entire system. Such a prediction would be almost impossible with quantum methods.

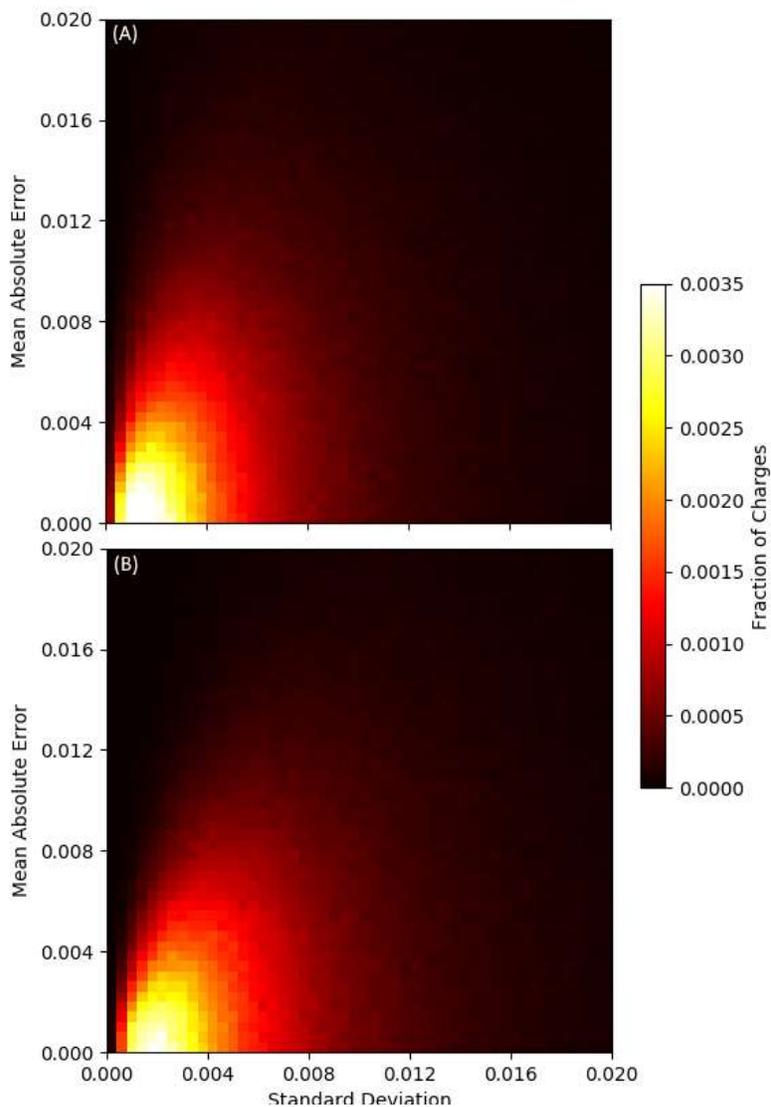

Figure 6: Ensembles of 4 (A) and 8 (B) networks, trained on GDB-5, predicting charges on GDB-7-9 structures. This demonstration shows that standard deviation between an ensemble of neural networks can be used to limit the maximum absolute error of the ensemble. The larger number of neural networks makes this effect more prominent.

**Error Estimation**

The ability of ML to determine the uncertainty of a prediction through schemes like Query by Committee is extremely powerful because it allows for a minimum number of reference calculations to be run when constructing optimized training sets. Although active learning been studied in the context of molecular energy modeling,[4, 28, 50-54] this concept has yet to be applied to charge prediction. Here we demonstrate how the consensus of an ensemble of neural networks can be used as a proxy for their accuracy in a "Query by Committee" framework (see earlier discussion in "Databases" section).[55]

To test this for charges, two ensembles (of size 4 and 8) of R=6.0 HIP-NN were trained to the GDB-5 database. The networks were then applied to predict charges in GDB-7-9. For each atomic charge, the ensemble average prediction error and ensemble prediction standard deviation was computed.

Figure 6 depicts density maps of prediction error vs ensemble standard deviation for the GDB-7-9 dataset for each ensemble. The standard deviation provides a typical bound for the ensemble error; the upper left quadrant of the density map is empty, demonstrating that there are no points with low ensemble standard deviation and high error. Points in the upper-left quadrant would be problematic for QBC – they correspond to points which are predicted confidently but inaccurately by the ensemble.

**Conclusion**

In this study, we have demonstrated the ability of HIP-NN to learn and predict several charge partitioning schemes with accuracy comparable to or better than basis set error of roughly 0.01 $e^-$. Our best performing R=8.0 network is capable of predicting charge with a MAE of less than 0.005e on 5 out of 6 in the COMP6 benchmarks. Additionally, the computational cost of ML charge prediction is roughly four orders of magnitude faster than DFT calculations for small molecules. We find that MSK charges, which are constrained to replicate the quantum molecular dipole, pose more difficulty for ML than do local charge schemes.

We demonstrate two applications of our ML charge models. First, we use a trained general-purpose network to calculate the IR spectrum of four small molecules. The predicted IR spectra is in excellent agreement with analogous spectrum from QM charge assignment. This demonstrates the transferability of charge prediction over a variety of systems without specifically tuning the network – it is proof of principle that a universal ML charge predictor is possible. Second, we generate charges for a large protein, where the application of quantum mechanics is extremely difficult. The speed of this prediction is evidence that ML methods can generate dynamic charges for fast and accurate electrostatic calculations in MD.

Additionally, we used the concept of *query-by-committee* to show a correlation between the accuracy and standard deviation of predictions made by an ensemble of ML models; when the ensemble agrees on a charge prediction, the prediction is more likely to be accurate. This is critical to the advancement of ML as applied to chemistry, as it facilitates active learning for the development of strong, diverse training sets. This is also useful when making predictions for unexplored systems, as the confidence for any ML prediction may be obtained without reference QM calculations.

The strong performance on COMP6 benchmarks and other applications demonstrates transferability of our ML approach to systems larger and more diverse than in the training set. Future models need not scale their data to sizes of target molecular systems. Datasets can be built using fragmentation and/or sampling of small molecules, and uncertainty can be estimated with query-by-committee. This provides a path forward to estimating chemical properties of systems intractable by QM methods.

Despite our success in training HIP-NN to various charge schemes, the underlying QM charge partitioning itself has drawbacks. The full charge density cannot be exactly represented within any charge assignment scheme. Two consequences are that different schemes disagree with each other, and that charge-derived quantities (such as the molecular dipole moment) may not be faithfully represented by atomic charge assignment. Future work may focus on overcoming the limitations of charge schemes by focusing instead on predictions of experimentally accessible quantities. For example, one may seek to learn a local charge model that best reconstructs the quantum molecular dipole. This, along with many other possible applications of ML to quantum chemical properties (such as bond orders, vibrational and electronic excitations) indicate tremendous potential present in the intersection of quantum chemistry and machine learning.

## Conflicts of interest

The authors are unaware of any conflicts of interest.

## Acknowledgements

The authors acknowledge support of the U.S. Department of Energy through the Los Alamos National Laboratory (LANL) LDRD Program. LANL is operated by Los Alamos National Security, LLC, for the National Nuclear Security Administration of the U.S. Department of Energy under contract DE-AC52- 06NA25396. This work was done in part at the Center for Nonlinear Studies (CNLS) and the Center for Integrated Nanotechnologies (CINT) at LANL.  We also acknowledge the LANL Institutional Computing (IC) program and ACL data team for providing computational resources.

## Notes and references